# Towards Conceptual Multidimensional Design in Decision Support Systems

Olivier Teste

Université Paul Sabatier - IRIT/SIG
118 Route de Narbonne - 31062 Toulouse cedex 04 (France)
e-mail: teste@irit.fr

**Abstract.** Multidimensional databases support efficiently on-line analytical processing (OLAP). In this paper, we depict a model dedicated to multidimensional databases. The approach we present designs decisional information through a constellation of facts and dimensions. Each dimension is possibly shared between several facts and it is organised according to multiple hierarchies. In addition, we define a comprehensive query algebra regrouping the more popular multidimensional operations in current commercial systems and research approaches. We introduce new operators dedicated to a constellation. Finally, we describe a prototype that allows managers to query constellations of facts, dimensions and multiple hierarchies.

## 1   Introduction

In order to improve decision-making process in companies, decision support systems are built from sources (operational databases). These dedicated systems are based on the data warehousing approach [4, 11, 24]. A data warehouse [11] stores large volumes of data, which are extracted from multiple, distributed, autonomous and heterogeneous data sources [4, 11, 24] and they are available for querying.

### 1.1 The Problem

In previous works, we specified a functional architecture of the decision support systems [18, 19], based on a dichotomy of two repositories; a data warehouse collects source data, which is relevant for the decision-makers, and it keeps data changes over the time whereas data marts are deduced from the data warehouse and they are dedicated to specific analyse (each data mart is subject-oriented). This architecture distinguishes several issues laying the foundation of our study (*cf.* figure 1).

- The **integration** generates a global source from data sources; it is virtual and it is described according to the ODMG data model. The motivating for using the object paradigm at the integration is that it has proven to be successful in complex data modelling [2].
- The **construction** generates a data warehouse as a materialised view [8] over the global source. It is not organised according to a multidimensional model [12]. We justify this choice by the fact that this modelling generates a lot of redundant data [4, 11, 12] limiting efficient warehouse management. We defined a flexible temporal object-oriented data warehouse model in [18, 19].

Due to manager requirements, we provide two approaches for improving the decision making process.

In the first approach the managers exploit the warehouse data to make global analyses. They are helped by database specialists who can directly query warehouse data using powerful and expressive languages. This approach has the advantage of allowing global analyses of the decisional information.

In the second approach the managers make themselves their analyses. They require advanced tools that facilitate analyses and multidimensional operations. We provide a solution based on two steps:

– The **organisation** models data for supporting efficiently OLAP ("On-Line Analytical Processing") applications [5] in several subject-oriented data marts. The data marts may be designed according to a multidimensional model [12, 17].
– The **interrogation** exploits decisional information. The managers improve their decisions through advanced tools facilitating OLAP applications.

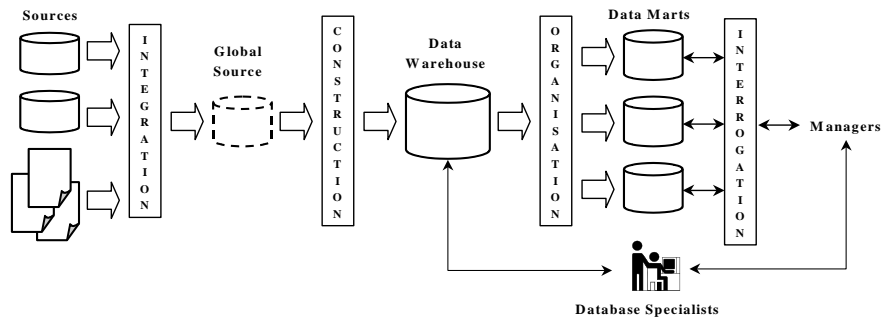

**Fig. 1.** Architecture of Decision Support Systems.

In this paper, we focus on this approach based on data mart generations where relevant data is stored "*multidimensionaly*". We depict a multidimensional model and we define a multidimensional query algebra.

### 1.2 Related Work

In academic research, multidimensional modelling has enjoyed spectacular growth [6]. One of the significant development is the proposal of the data cube operator [7]. Several approaches treat data as n-dimensional cubes where the data is divided in measures (facts) and dimensions [7, 9, 12], but the hierarchy between the parameters is not captured explicitly by the schema. Therefore, several proposals provide structured cube models, which capture dimension hierarchies [1, 3, 13, 14, 17]. Some models provide statistical objects where a structured hierarchy is related to an explicit aggregation function on a single measure supporting a set of queries [20]. To model dimensions of complex structures, several models were made in an object oriented framework [3, 16, 21]. Also, some proposals exploits the temporal nature of the multidimensional modelling [10, 15, 16].

Most of these proposals introduce constraints and specific modelling choices as ROLAP, MOLAP and OOLAP. Nevertheless, in [22] the authors provide a full conceptual approach through the *starER* model, which combines the star structure with the semantically rich constructs of the ER model. In the same way, the model we present is independent of the ROLAP, OOLAP or MOLAP context.

Moreover, existing approaches design a multidimensional database as a star schema [12]. This approach integrates only one fact. We argue that an extended multidimensional model in which a multidimensional database is designed as a constellation of facts and dimensions is a more efficient way for improving a powerful multidimensional modelling [17]. This extended model needs a query language integrating the more popular operations in current commercial systems and research approaches as well as some operations related to the constellation organisation. The main contribution of this paper is the comprehensive multidimensional query algebra that we define. We provide formal definitions of the most important multidimensional operations and we define two new operations related to the constellation organisation.

### 1.3 Paper Outline

Section 2 defines a multidimensional model supporting facts, shared dimensions and multiple hierarchies, independently of ROLAP, OOLAP or MOLAP contexts. Section 3 presents the query algebra related to the multidimensional model. Section 4 describes extensions of our prototype GEDOOH.

## 2 A Multidimensional Model

In the architecture that we depict in figure 1, a data mart is subject-oriented; it is dedicated to a specific class of users and it regroups all relevant information for supporting their decisional requirements. The data mart must be modelled "*multidimensionaly*" for improving analyses and decision making processes [12].

The multidimensional model we define is based on the idea of the "*constellation*" [17], in which data marts are composed of several facts and dimensions; each dimension is shared between facts and it can be associated to one or several hierarchies. Therefore, the managers can handle several facts according to shared dimensions, facilitating comparisons between several measures.

### 2.1 Facts

A fact reflects information that have to be analysed; for example, a factual data is the amount of sales occurring in shops.

**Definition 1**. A **fact** $F$ is defined by a tuple (*fname*, $M^{fname}$) where
– *fname* is a name,
– $M^{fname}=\{m_1, m_2,…, m_m\}$ is a set of attributes where each $m_k$ represents one measure.

### 2.2 Dimensions and Hierarchies

A dimension reflects information according to which data of facts will be analysed. A dimension is organised through parameters, which conform to one or several hierarchies; the dimensions of interest may be the shop location, the time,…

**Definition 2**. A **dimension** $D$ is defined by a tuple (*dname*, $A^{dname}$, $H^{dname}$) where
– *dname* is a name,
– $A^{dname}$ is a set of attributes,
– $H^{dname}=<H^{dname}_1, H^{dname}_2,…, H^{dname}_h>$ is an ordered set of hierarchies ($H^{dname}_1$ is called the current hierarchy).

The parameters are organised according to hierarchies. Within a dimension, values of different parameters are related through a family of roll up functions, denoted $\rho_{roll}$, according to each hierarchy defined on them. A roll up function $\rho_{roll}^{H(pj \to pj')}$ associates a value $v$ of a parameter $p_j$ with a value $v'$ of an upper parameter $p_j'$ in the hierarchy $H$.

**Definition 3**. A **hierarchy** $H^{dname}_i$ is defined by a tuple ($hname$, $P^{hname}$) where
- $hname$ is a name,
- $P^{hname}=<p_{i1}, p_{i2},..., p_{hi}>$ is an ordered set of parameters where $\forall j \in [i_1..h_i]$, $p_j \in A^{dname}$.

Note that $A^{dname}$ contains a distinguished parameter *all*, such that $dom(all)=\{All\}$. This attribute defines the upper granularity of hierarchies; for every hierarchy $H^{dname}_j \in H^{dname}$, $H^{dname}_j =< p_{i1}, p_{i2},..., all >$.

### 2.3 Constellation Schema

A data mart is modelled according to a constellation schema; it is composed of several facts and several dimensions, which are possibly shared between facts.

**Definition 4**. A **constellation schema** $S^{DM}$ is defined by a tuple ($sname$, $FACT$, $DIM$, $Param^{sname}$) where
- $sname$ is a name,
- $FACT=<F_1, F_2,..., F_u>$ is an ordered set of facts ($F_1$ is called the current fact),
- $DIM=\{D_1, D_2,..., D_v\}$ is a set of dimensions,
- $Param^{sname}: FACT \to 2^{DIM}$ is a function such that $Param^{sname}(F_i)=<D_i^1, D_i^2,..., D_i^{wi}>$. It returns an ordered set of dimensions which are associated to the fact $F_i$ (for the current fact $F_1$, $D_1^1$ and $D_1^2$ are called the current dimensions).

**Example.** The case we study is taken from commercial domain and it concerns shop channels. The data mart supports analyses about sales and purchases related to various commercial shop channels and commercial warehouses. Figure 2 represents the example of a data mart; we use extended relational notations: '⬚' represents a fact and '↯' represents a dimension. Note that the constellation is named "*channalyse*".

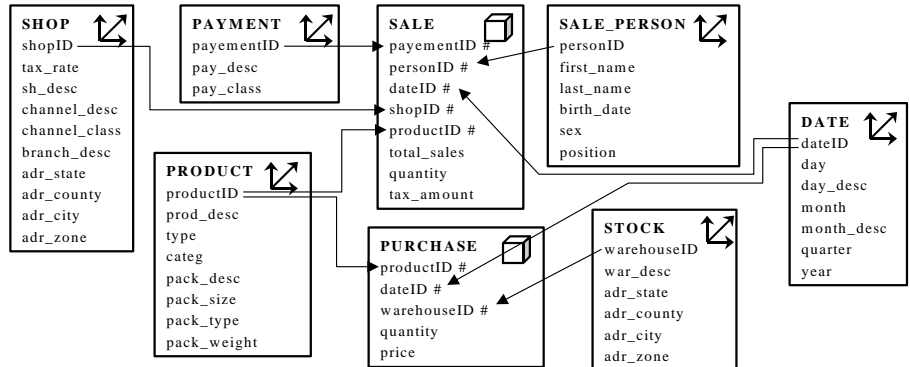

**Fig. 2.** Example of a Constellation Schema.

Along each dimension, the administrator defines one, or possibly several, hierarchy(ies). These hierarchies offer various views for analysed data; *e.g.* users can analyse sales according to dates and shops, and they can display analysed data with an administrative organisation of the country (*adr_state*, *adr_county*, *adr_city*) or with a specific organisation (*adr_city*, *adr_zone*). The hierarchies are defined as follows.

- $H^{shop}_1$ = ("h_shop_channel", <shopID, channel_class, branch_desc, all>)
- $H^{shop}_2$ = ("h_shop_administrative", <shopID, city, county, state, all>)
- $H^{shop}_3$ = ("h_shop_zone", <shopID, city, zone, all>)
- $H^{payment}_1$ = ("h_payment", <paymentID, pay_class, all>)
- $H^{person}_1$ = ("h_person_position", <personID, position, all>)
- $H^{product}_1$ = ("h_product_category", <prodID, type, categ, all>)
- $H^{date}_1$ = ("h_date_gregorian", <dateID, day, month, quarter, year, all>)
- $H^{stock}_1$ = ("h_stock_administrative", <warehouseID, city, county, state, all>)
- $H^{stock}_2$ = ("h_stock_zone", <warehouseID, city, zone, all>)

## 3 A Comprehensive Multidimensional Query Algebra

Here, we express in a query algebra the most popular OLAP operators introduced in the scientific literature and we provide new operators related to the constellation organisation.

### 3.1 Data Displaying: "n-table"

A constellation is displayed within an n-table according to columns, rows and planes. The current fact $F_1$ is used to define the displayed plane. The current dimensions $D_1^1$ and $D_1^2$ of the current fact define displayed lines and rows. For each current dimension, the upper level is displayed according to the current hierarchy. Note that because of the constellation feature, we do not display the complete information stored in data marts; more precisely, only the measures of the current fact are displayed according to the current dimensions and their current hierarchies.

**Example.** We deal with the previous example. The current fact is "*sale*" and the current dimensions are "*shop*" and "*payment*" displayed according to the hierarchies "*h_shop_channel*" and "*h_payment*".

**Table 1.** Example of a Constellation Displaying.

| Sale | | Shop / $H^{shop}_1$ | | | | |
|---|---|---|---|---|---|---|
| | | branch_desc | BR1 | BR2 | BR3 | BR4 |
| Payment / $H^{payment}_1$ | pay_class | total_sales, tax_amount, quantity | | | | |
| | PC1 | | (58,6, 2) | (67,7, 3) | (58,6, 1) | (68,7, 2) |
| | PC2 | | (60,6, 3) | (55,6, 3) | (50,5, 1) | (65,7, 3) |
| | PC3 | | (45,5, 1) | (50,5, 1) | (52,5, 1) | (64,6, 2) |
| Sale_person.position="manager" | | | | | | |
| Product.categ="C1" | | | | | | |
| Date.year=2000 | | | | | | |

### 3.2 Multidimensional Operations

We first define relational operators in the multidimensional algebra; we adopt the most popular operators (Join, Aggregate, Union, Intersect, and Difference). The operation Slice and Dice is used on a dimension and it removes values of the dimension that do not satisfy a restricted condition. Note that this operator realises selecting (or restricting) in relational terminology.

Because the complete information stored in data marts is not displayed, we define rotate operators for displaying measures according to various parameters. We adopt rotate operators introduced in [1], and we define a new rotation between facts.

**Definition 5**. The **DRotate** operation permutes two dimensions $D_i$ and $D_j$ of a fact $F$. $DRotate(Sh, F, D_i, D_j)=Sh'$ where
- $Sh=(sname, FACT, DIM, Param^{sname})$ is a constellation schema,
- $F \in FACT$ is a fact,
- $D_i \in DIM$ and $D_j \in DIM$ are two dimensions | $Param(F)=<\ldots, D_i, \ldots, D_j, \ldots>$.
$Sh'=(sname, FACT, DIM, Param^{sname'})$ where $Param^{sname'}(F)=<\ldots, D_j, \ldots, D_i, \ldots>$ and $\forall F \in FAI, F_k \neq F, Param^{sname'}(F_k)=Param^{sname}(F_k)$.

**Definition 6**. The **HRotate** operation permutes two hierarchies $H^{dname}_i$ and $H^{dname}_j$ of a dimension $D$. $HRotate(Sh, D, H^{dname}_i, H^{dname}_j)=Sh'$ where
- $Sh=(sname, FACT, DIM, Param^{sname})$ is a constellation schema,
- $D \in DIM$ is a dimension,
- $H^{dname}_i \in H^{dname}$ and $H^{dname}_j \in H^{dname}$ are two hierarchies | $H^{dname}=<\ldots, H^{dname}_i, \ldots, H^{dname}_j, \ldots>$.
$Sh'=(sname, FACT, DIM', Param^{sname'})$ where
- $DIM'=DIM-\{D_i\}+\{D_i'\}$ | $D_i'=(dname_i, P^{dname i}, H^{dname i}=<\ldots, H^{dname}_j, \ldots, H^{dname}_i, \ldots>)$
- $\forall D \in FACT$, if $D_i \in Param^{sname}(F)$ then $Param^{sname'}(F)=Param^{sname}(F)-\{D_i\}+\{D_i'\}$ else $Param^{sname'}(F)=Param^{sname}(F)$.

**Definition 7**. The **FRotate** operation permutes two facts $F_i$ and $F_j$ of a constellation schema $Sh$. $FRotate(Sh, F_i, F_j)=Sh'$ where
- $Sh=(sname, FACT, DIM, Param^{sname})$ is a constellation schema,
- $F_i \in FACT$ and $F_j \in FACT$ are two facts | $FACT=<\ldots, F_i, \ldots, F_j, \ldots>$.
$Sh'=(sname, FACT', DIM, Param^{sname})$ where $FACT'=<\ldots, F_j, \ldots, F_i, \ldots>$.

**Example.** We complete the previous example. Managers change the dimensions in order to analyse measures according to other parameters. They permute "*Shop*" and "*Date*" as well as "*Payment*" and "*Product*". `DRotate(DRotate("channalyse", Payment, Product), Shop, Date)`

**Table 2.** N-table Representing the Constellation after Rotations.

| Sale | | Date / $H^{date}_1$ | | | |
|---|---|---|---|---|---|
| | | year | 1998 | 1999 | 2000 |
| Product / $H^{product}_1$ | categ | total_sales, tax_amount, quantity | | | |
| | C1 | | (58,6, 2) | (67,7, 3) | (58,6, 1) |
| | C2 | | (60,6, 3) | (55,6, 3) | (50,5, 1) |
| | C3 | | (45,5, 1) | (50,5, 1) | (52,5, 1) |
| **Sale_person**.position="manager" | | | | | |
| **Payment**.pay_class ="PC1" | | | | | |
| **Shop**.branch_class="BR1" | | | | | |

The positions (values) of each parameter are ordered. We introduce one operator for changing these positions.

**Definition 8**. The operation **Switch** permutes two positions (values) $pos_{j1}$ and $pos_{j2}$ of a parameter $p$. $Switch(Sh, d, p, pos_{j1}, pos_{j2})=Sh'$ where
- $Sh=(sname, FACT, DIM, Param^{sname})$ is a constellation schema,
- $D \in DIM$ is a dimension,
- $p \in P^{dname}$ is a parameter of the current hierarchy | $p \in H^{dname}_1$,
- $pos_{j1} \in dom(p)$ and $pos_{j2} \in dom(p)$ are two positions (values) of the parameter $p$.
$Sh'$ is the result where $pos_{j1}$ and $pos_{j2}$ are permuted in the hierarchy $H^{dname}_1$.

The RollUp and DrillDown operations are probably the most important operations for OLAP; they allow users to change data granularities.

**Definition 9**. The **DrillDown** operation inserts into the current hierarchy of a dimension $D_i$, a parameter $p_j$ at a lower granularity. $DrillDown(Sh, D_i, p_j)=Sh'$ where
- $Sh=(sname, FACT, DIM, Param^{sname})$ is a constellation schema,
- $D_i \in DIM$ is a dimension such that $D_i=(dname_i, P^{dnamei}, H^{dnamei})$,
- $p_j \in P^{dnamei}$ is a parameter (it will be integrated in the current hierarchy of $D_i$).

$Sh'=(sname, FACT, DIM', Param^{sname'})$ where
- $DIM'=DIM-\{D_i\}+\{D_i'\}$ | $D_i'=(dname_i, P^{dnamei}+\{p_j\}, H^{dnamei'}=<<p_j>+H^{dname}{}_1, H^{dname}{}_2,\ldots, H^{dname}{}_h>)$ and
- $\forall F \in FACT$, if $D_i \in Param^{sname}(F)$ then $Param^{sname'}(F)=Param^{sname}(F)-\{D_i\}+\{D_i'\}$ else $Param^{sname'}(F)=Param^{sname}(F)$.

**Definition 10**. The **RollUp** operation inserts into the current hierarchy of a dimension $D_i$, a parameter $p_j$ corresponding to an upper granularity. $RollUp(Sh, D_i, p_j)=Sh'$ where
- $Sh=(sname, FACT, DIM, Param^{sname})$ is a constellation schema,
- $D_i \in DIM$ is a dimension | $D_i=(dname_i, P^{dnamei}, H^{dnamei})$,
- $p_j \in parameters^{dnamei}$ is a parameter, which will be integrated in the current hierarchy of the dimension $D_i$.

$Sh'=(sname, FACT, DIM', Param^{sname'})$ where
- $DIM'=DIM-\{D_i\}+\{D_i'\}$ | $D_i'=(dname_i, P^{dnamei}+\{p_j\}, H^{dnamei'}=<H^{dname}{}_1+<p_j>, H^{dname}{}_2,\ldots, H^{dname}{}_h>)$ and
- $\forall F \in FACT$, if $D_i \in Param^{sname}(F)$ then $Param^{sname'}(F)=Param^{sname}(F)-\{D_i\}+\{D_i'\}$ else $Param^{sname'}(F)=Param^{sname}(F)$.

In order to ease analyses, in [1] the authors introduce operations allowing an uniform treatment of parameters and measures; one operator converts parameters into measures and another one creates parameters from specified measures. We adopt these operations in the constellation framework.

**Definition 11**. The **Push** operation converts parameters into measures. $Push(Sh, d, p, f)=Sh'$ where
- $Sh=(sname, FACT, DIM, Param^{sname})$ is a constellation schema,
- $D \in DIM$ is a dimension,
- $p \in P^{dname}$ is a parameter of the dimension $D$,
- $F \in FACT$ is a fact | $D \in Param(F)$.

$Sh'=(sname, FACT', DIM', Param^{sname'})$ where
- $FACT'=FACT-\{F\}+\{F'\}$ | $F'=(fname, M^{fname}+\{p\})$,
- $DIM'=DIM-\{D_i\}+\{D_i'\}$ | $P^{dname'}=P^{dname}-\{p\}$ and
- $Param^{sname'}(F') = Param^{sname}(F)-\{D_i\}+\{D_i'\}$, $\forall F'' \in FACT, F'' \neq F'$, $Param^{sname'}(F'') = Param^{sname}(F)$.

**Definition 12**. The **Pull** operation converts measures into parameters. $Pull(Sh, F, m, D)=Sh'$ where
- $Sh=(sname, FACT, DIM, Param^{sname})$ is a constellation schema,
- $F \in FACT$ is a fact,
- $m \in M^{fname}$ is a measure of the fact $F$,

- $D \in DIM$ is a dimension | $D \in Param(F)$.

$Sh'=(sname, FACT', DIM', Param^{sname\prime})$ where
- $FACT'=FACT-\{F\}+\{F'\} \mid F'=(fname, M^{fname}-\{m\})$,
- $DIM'=DIM-\{D_i\}+\{D_i'\} \mid P^{dname\prime}=P^{dname}+\{m\}$ and
- $Param^{sname\prime}(F') = Param^{sname}(F)-\{D_i\}+\{D_i'\}$, $\forall F'' \in FACT$, $F'' \neq F'$, $Param^{sname\prime}(F'') = Param^{sname}(F)$.

To convert a constellation into several star schemas (constellations composed of one fact), we introduce two operators. They allow users to reduce schemas.

**Definition 13**. The **TSplit** operation generates several sub schemas from a constellation schema according to its facts. Each generated schema is composed of one fact. $TSplit(Sh)=\{Sh_1,\ldots, Sh_u\}$ where
- $Sh=(sname, FACT, DIM, Param^{sname})$ is a constellation schema.

$\forall i \in [1..u]$, $Sh_i=(sname, FACT', DIM', Param^{sname\prime})$ is a resulting sub schema. Its a constellation schema composed of one fact such that $FACT'=\{F_i\}$, $DIM'=\{D \mid D \in DIM \wedge D \in Param(F_i)\}$ and $Param'(F_i)=Param(F_i)$.

**Definition 14**. The **Split** operation generates several sub schemas from a constellation schema, which is composed of one fact. Each generated sub schema results from a selection. $Split(Sh, D, p)=\{Sh_1, Sh_2,\ldots, Sh_s\}$ where
- $Sh=(sname, FACT, DIM, Param^{sname})$ is a constellation schema.
- $D \in DIM$ is a dimension,
- $p \in P^{dname}$ is a parameter of the dimension $D \mid dom(p)=\{pos_1, pos_2,\ldots pos_s\}$.

$\forall i \in [1..n]$, $Sh_i=(sname, FACT, DIM, Param^{sname})$ is a resulting sub schema according to the slice operation $Slice(Sh, D, pred(pos_i))$.

## 4 Implementation

In previous works, we have implemented a prototype allowing administrators both to define and to generate data warehouses and data marts. This prototype is called **GEDOOH**. It is based on three components: a graphical interface, an automatic data warehouse generator, and an automatic data mart generator.

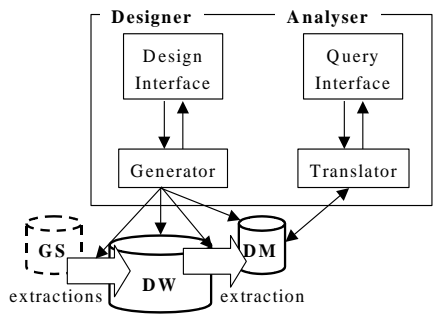

**Fig. 3.** GEDOOH Architecture.

GEDOOH helps administrators in designing data warehouses and data marts. It is based on extended UML notations for displaying schemas.
- Firstly, the administrator defines a data warehouse from a graph of the global source (or a data mart from a graph of the data warehouse).
- Secondly, the generators create automatically the data warehouse (or the data mart) according to the graphical definitions. The schema, the first extraction (which populates the data warehouse or the data mart) and the refresh process are generated.

This tool is implemented in Java (jdk 1.3) on top of a relational database management system (Oracle) and it is operational; its source code represents approximately 8000 lines of Java code.

Now, we are implementing extensions in order to validate the model we present in this paper and its associated query algebra. We add a user component allowing the managers both to display and to query constellation schemas of the generated data marts. The extension (the user component) is composed of two parts: an interface and a query translator.

- The **query interface** displays an n-table representing a constellation. This component uses internal structures of the displayed information. Each multidimensional operation is treated by the query translator component.
- The **query translator** translates each multidimensional operation in a relational query. This component sends the relational query to the RDBMS and it translates the result in internal structures.

**Fig. 4.** Example of a Constellation Displaying through the GEDOOH Query Interface.

## 5  Conclusion

We first introduce an architecture of decision support systems distinguishing several issues and laying the foundation for our study. Based on the architecture, this paper deals with the data mart designing and querying.

The multidimensional model we define is based on the idea of the "*constellation*", in which data marts are composed of several facts and dimensions; each dimension is shared between facts and it can be associated to one or several hierarchies. Shared dimensions facilitates comparisons between several measures according to the same dimensional data organisation (same hierarchies, same parameters…). This approach provides a unified framework for the multidimensional modelling independently of the ROLAP, OOLAP or MOLAP context. We develop a query algebra for the data marts. We express in a comprehensive algebra the most popular OLAP operators and we provide new operators related to the constellation organisation (FRotate, TSplit).

We are currently working on extending the tool GEDOOH. It allows administrators to generate a data warehouse from sources through a graphical interface. We have extended algorithms for generating data marts from data warehouses. Now, we develop solutions for querying data marts based on the query algebra.

We will investigate meta-modelling issues and we plan to develop a method for designing decision support systems. We must provide a design method; it must be composed of models (with concepts and constraints), a complete process and a tool.